\documentclass{article}
\makeatletter
\@addtoreset{equation}{section}

\makeatother

\begin{document}

\begin{flushright}
Feb 2003

KEK-TH-867
\end{flushright}

\begin{center}

\vspace{5cm}

{\Large Target Space Approach to Closed String Tachyons}

\vspace{2cm}

Takao Suyama \footnote{e-mail address : tsuyama@post.kek.jp}

\vspace{1cm}

{\it Theory Group, KEK}

{\it Tsukuba, Ibaraki 305-0801, Japan}

\vspace{4cm}

{\bf Abstract} 

\end{center}

We use the low energy effective theory of string theory to investigate condensations of closed string tachyons 
propagating in the bulk. 
The c-function is related to the total energy of the system via the effective action. 
A possible modification of the c-theorem is discussed. 
We also deduce endpoints of the decays by investigating scalar potential of gauged supergravities. 
A string theory corresponding to a condensation is argued. 

\newpage

\vspace{1cm}

\section{Introduction}

\vspace{5mm}

The decay of an unstable vacuum is a dynamical process. 
To study this, one has to consider non-supersymmetric theories. 
Therefore, investigations of this kind of phenomena are difficult, in general. 

In this paper, our interest is on condensations of tachyons which can propagate in the bulk. 
What we have to clarify would be the following two questions. 
The first one is into what theory the original tachyonic theory decays, and the second one is how to analyze the 
decay. 
The answer to the latter is, in principle, clear; one can use a closed string field theory \cite{closedSFT}, or 
matrix 
formulations of string theories and M-theory \cite{BFSS}\cite{IKKT}\cite{DVV}. 
However, such an analysis will be, in practice, difficult to perform. 
The former question is more difficult to answer. 
There are two naive expectations for endpoints of the decay. 
One possibility is to decay into a non-critical string theory, since the Zamolodchikov's c-theorem 
\cite{Zamolodchikov}  states that 
the central charge of a worldsheet theory decreases via a relevant deformation. 
The other is discussed in our previous paper \cite{suyama}. 
We argued that there may be a process which connects two critical string theories via a tachyon condensation, 
when the 
corresponding low energy effective theory has a scalar potential, two of whose critical points correspond to these 
critical string theories. 
We also argued that since the theory flows into a strong coupling theory, the worldsheet analysis, i.e. arguments 
based on the c-theorem, might be irrelevant for the dynamics of string theory. 

In this paper, we will employ the low energy effective theory as a tool, and try to answer the two questions 
mentioned above. 
The results of our investigations are summarized as follows. 
The Zamolodchikov's c-function of a sigma model is equal to the action functional of the corresponding low energy 
effective theory \cite{Tseytlin}. 
This expression of the c-function was used recently \cite{GHMS} to show that worldsheet renormalization group 
flows of a kind correspond to processes which are favored energetically. 
The relation between the c-function and the effective action can be naturally generalized, and as a result, 
the decrease of the c-function is related to the decrease of the total energy of the spacetime theory, when the 
dilaton is trivial. 
Otherwise, the relation between the c-function and the energy does not necessarily hold. 
Therefore, we would like to claim that the c-theorem cannot always be a guide to deduce endpoints of the 
decay. 
We further study the final states of the decay by using the low energy effective theory which can be regarded as 
the zeroth-order approximation of the closed string field theory. 
In some cases, one can relate an evolution of the tachyon vev to a deformation of the geometry of a manifold on 
which string theory or M-theory are compactified. 

As we will show in this paper, the use of the effective theory, in particular a gauged supergravity, would be very 
fruitful for the investigations of bulk tachyon condensations. 
We can obtain some explicit results since some of the effective theories have been studied exhaustively. 
More detailed information of other gauged supergravities, in particular a relation to string theory and M-theory, 
will be valuable for the study of tachyon condensations. 

This paper is organized as follows. 
In section \ref{Ddep}, we review the results of \cite{suyama}. 
In section \ref{Dindep} we discuss a tachyon condensation different from those in \cite{suyama}, in which the 
worldsheet analysis would not break down. 
We suggest here that there is a subtlety on the comparison of central charges when the background changes during 
the process. 
In section \ref{center}, we generalize the expression of \cite{Tseytlin}. 
We then argue how it is related to the energy and how the decrease of the 
c-function can deviate from the decrease of the spacetime energy. 
In section \ref{potential}, we deduce the possible endpoints of bulk tachyon condensations based on the analysis 
of the scalar potentials of the low energy effective theories. 
Section \ref{discussion} is devoted to discussion.

\vspace{1cm}

\section{Worldsheet RG}  \label{Ddep}

\vspace{5mm}

As we have learned from the study of condensations of open string tachyons \cite{open} and localized 
closed string tachyons \cite{localized}, there is a prescription to deduce an endpoint of the tachyon 
condensations. 
Suppose that there is a string theory with a tachyon in its mass spectrum. 
Then a condensation of the tachyon can be described by adding the corresponding vertex operator to 
the worldsheet action. 
The added operator is usually taken to be a zero-momentum tachyon vertex and it is a relevant 
operator. 
The condensation would make a change of the worldsheet theory to cure the instability indicated by 
the presence of the tachyon. 
The endpoint of the condensation is believed to be described by a theory which is the IR fixed point 
of the RG flow induced by the relevant operator. 

It is tempting to speculate that this prescription would also apply to condensations of tachyons 
propagating in the bulk. 
However, there seems to be a puzzle concerning the central charge of the worldsheet theory. 

\vspace{3mm}

The tree level dynamics of strings in a non-trivial background is described by a non-linear sigma 
model. 
The background field configuration is encoded in coupling functions of the sigma model. 
For the sigma model to be conformal, beta-functionals of the coupling functions must vanish, and this 
condition is equivalent to the on-shell condition of the background fields. 
Since one imposes the conformal invariance, the corresponding string theory is critical, that is, its 
central charge is zero after adding contributions from the ghosts. 

Assume that a string theory in a background contains a tachyon propagating in the bulk. 
The presence of the tachyon would be a signal of an instability of the background, and the theory 
would be deformed so as to cure such the instability. 
This process would be realized via a condensation of the tachyon, as in the case of open string 
tachyons \cite{open} and of localized closed string tachyons \cite{localized}. 
(See also \cite{70s} for earlier study on this phenomena.)
During the condensation of localized closed string tachyons, it is claimed that the central charge 
does not decrease, due to the fact that the relevant space for the tachyons has an infinite volume 
\cite{HKMM}. 
This argument does not apply to situations in which tachyons come from a compact space. 
Also in the case of bulk tachyons, one cannot conclude that the central charge does not need to 
decrease. 
Therefore, according to the above discussions, one might expect that a condensation of a bulk tachyon 
would lead the theory to a non-critical string theory. 

\vspace{3mm}

On the other hand, the same situation could be investigated by using the spacetime effective theory 
\cite{suyama}. 
Suppose that there is a non-linear sigma model, the bosonic part of whose low energy effective action is 
\begin{eqnarray}
S &=& \frac1{2\kappa^2}\int d^Dx\sqrt{-g}e^{-2\Phi}\left[ R+4\partial_\mu\Phi\partial^\mu\Phi
     -\frac1{12}H_{\mu\nu\rho}H^{\mu\nu\rho}\right. \nonumber \\
  & & \hspace*{2cm}\left.\frac{ }{ }
     -f_{ab}(\phi)F^a_{\mu\nu}F^{b\mu\nu}-g_{IJ}(\phi)D_\mu\phi^I D^\mu\phi^J-V(\phi)\right]. 
          \label{gaugedSUGRA}
\end{eqnarray}
The action of this form appears, for example, in the heterotic string theory compactified on a torus 
with flux \cite{hetero}. 
Suppose that the scalar potential has an unstable critical point $\phi^I=\phi^I_1$. 
If there is another critical point $\phi^I=\phi^I_2$ which is less unstable than $\phi^I=\phi^I_1$ and 
$V(\phi_1)>V(\phi_2)$, then it would be natural to expect that the former vacuum decays into the latter vacuum via 
a condensation of scalar fields, i.e., a shift of the vevs. 
This process would be interpreted as a bulk tachyon condensation. 
However, both vacua $\phi^I=\phi^I_1,\phi^I_2$, with appropriate backgrounds, are classical solutions of the 
effective theory (\ref{gaugedSUGRA}). 
Therefore, each vacuum defines a critical string theory with vanishing central charge, and thus the central charge 
does not decrease during the process. 
Note that $\alpha'$-corrections could be controlled when the effective theory (\ref{gaugedSUGRA}) possesses some 
amount of spacetime supersymmetry. 

\vspace{3mm}

A possible resolution of this puzzle was proposed in \cite{suyama}. 
It is summarized as follows. 
The decay discussed above indeed decrease the central charge, which can be shown in the range of validity of the 
tree level approximation, but whole trajectory of the decay process in the theory space does not lie within the 
range, and comparison of central charges of initial and final theories is not possible. 

\vspace{3mm}

Let us explain our claim mentioned above. 
To be definite, we consider Type II string theory. 
The spacetime effective action of the NS-NS fields $({\cal G}_{MN},{\cal B}_{MN},\Phi)$ is 
\begin{equation}
S = \frac1{2\kappa^2}\int d^{10}x\sqrt{-{\cal G}}e^{-2\Phi}\left[
    {\cal R}+4(\hat{\nabla}\Phi)^2-\frac1{12}{\cal H}_{LMN}{\cal H}^{LMN} \right].
      \label{10dim}
\end{equation}
Here ${\cal R},\hat{\nabla}_M$ are the scalar curvature and the covariant derivative with respect 
to the metric ${\cal G}_{MN}$, and ${\cal H}_{LMN}$ is the field strength of ${\cal B}_{MN}$. 

One can obtain a low energy effective action in $d$ dimensions via a compactification. 
In the ordinary toroidal compactifications, the reduction is performed by constructing the fields in terms of 
fields in $d$ dimensions 
\begin{equation}
{\cal G}_{MN} = \left[ 
\begin{array}{cc}
g_{\mu\nu}+A_\mu^lA_{\nu l} & A_{\mu m} \\ A_{\nu n} & G_{mn}
\end{array}
\right], \hspace{5mm} \mbox{etc.}
   \label{toroidal}
\end{equation}
and substituting them into the action (\ref{10dim}). 
Here $\mu,\nu=0,\cdots,d-1$ label non-compact directions and $m,n=d,\cdots,9$ label compact (internal) directions. 
In the toroidal compactifications, $g_{\mu\nu}$ etc. are taken to be independent of the internal coordinates $y^m$. 

One can consider a more general compactification by allowing some $d$-dimensional fields to be $y^m$-dependent, 
and such a reduction will produce a non-trivial scalar potential in the reduced action. 
This is known as the Scherk-Schwarz reduction \cite{Scherk-Schwarz}. 
We will derive the reduced action below. 
For simplicity, we will omit terms including gauge fields $A_{\mu m}$ in $d$ dimensions. 
The resulting action is meaningful as long as we consider classical solutions with constant scalars, i.e., vacuum 
solutions. 

The reduction ansatz we make is 
\begin{equation}
{\cal G}_{MN} = \left[ 
\begin{array}{cc}
g_{\mu\nu}(x) & 0 \\ 0 & G_{mn}(x,y)
\end{array}
\right], \hspace{5mm} {\cal B}_{MN} = \left[ 
\begin{array}{cc}
B_{\mu\nu}(x) & 0 \\ 0 & B_{mn}(x,y)
\end{array}
\right], \hspace{5mm} \Phi = \Phi(x)
    \label{ansatz}
\end{equation}
By substituting this ansatz into (\ref{10dim}), one obtains 
\begin{eqnarray}
S &=& \frac1{2\kappa^2}\int d^{10}x\sqrt{-g}e^{-2\phi}\left[ 
R+4(\nabla\phi)^2-\frac1{12}H_{\mu\nu\rho}H^{\mu\nu\rho} \right. \nonumber \\
& & \hspace*{5mm} \left. +\frac14\nabla G^{mn}\cdot\nabla G_{mn}
    -\frac14G^{km}G^{ln}\nabla B_{kl}\cdot\nabla B_{mn}-V \right], 
     \label{twisted_action}
\end{eqnarray}
where 
\begin{eqnarray}
\phi &=& \Phi-\frac14\log\det G, \\
V &=& -G^{mn}R_{mn}+\frac1{12}G^{il}G^{jm}G^{kn}H_{ijk}H_{lmn}. 
\end{eqnarray}
Here $R,\nabla_\mu$ are the scalar curvature and the covariant derivative with respect to the metric $g_{\mu\nu}$, 
and $R_{mn}$ is the Ricci tensor with respect to $G_{mn}$. 
If the integrand is independent of $y^m$, the action (\ref{twisted_action}) can be regarded as an effective action 
in $d$ dimensions with a suitable rescaling of $\kappa$. 

For later purpose, we show the equations of motion of $G_{mn},B_{mn}$, 
\begin{eqnarray}
R_{mn}+\cdots &=& 0, \\
G^{kl}\nabla_kH_{lmn}+\cdots &=& 0,
\end{eqnarray}
where $\cdots$ denotes terms including $\nabla_\mu G_{mn},\nabla_\mu B_{mn}$. 
In terms of $d$-dimensional fields, these equations would correspond to 
\begin{equation}
\frac{\partial V}{\partial \phi^I}=0,
\end{equation}
for constant solutions, where $\phi^I$ represent $G_{mn},B_{mn}$ collectively, as long as the reduction is a 
consistent truncation. 

\vspace{3mm}

Since the relation between the ten-dimensional action (\ref{10dim}) and the reduced action (\ref{twisted_action}) 
is clear, 
one can easily obtain one-loop beta-functionals of a sigma model whose low energy effective action is the reduced 
action (\ref{twisted_action}), from the beta-functionals \cite{CFMP}
\begin{eqnarray}
\beta^{\cal G}_{MN} &=& {\cal R}_{MN}+2\hat{\nabla}_M\hat{\nabla}_N\Phi-\frac14{\cal H}_{MKL}{{\cal H}_N}^{KL}, \\
\beta^{\cal B}_{MN} &=&-\frac12\hat{\nabla}^L{\cal H}_{LMN}+\hat{\nabla}^L\Phi\ {\cal H}_{LMN}, \\
\beta^\Phi &=&-\frac12\hat{\nabla}^2\Phi+(\hat{\nabla}\Phi)^2-\frac1{24}{\cal H}_{LMN}{\cal H}^{LMN},
\end{eqnarray}
by substituting the ansatz (\ref{ansatz}). 
We assume, for simplicity, that $G_{mn},B_{mn}$ are independent of $x^\mu$. 
Then one obtains 
\begin{eqnarray}
\beta^{\cal G}_{\mu\nu} &=& R_{\mu\nu}+2\nabla_\mu\nabla_\nu\Phi-\frac14H_{\mu\rho\lambda}{H_\nu}^{\rho\lambda}, \\
\beta^{\cal G}_{mn} &=& R_{mn}-\frac14G^{ik}G^{jl}H_{mij}H_{nkl}, \\
\beta^{\cal B}_{\mu\nu} &=& -\frac12\nabla^\rho H_{\rho\mu\nu}+\nabla^\rho\Phi\ H_{\rho\mu\nu}, \\
\beta^{\cal B}_{mn} &=& -\frac12G^{kl}\nabla_kH_{lmn}, \\
\beta^\Phi &=& -\frac12\nabla^2\Phi+(\nabla\Phi)^2-\frac1{24}G^{il}G^{jm}G^{kn}H_{ijk}H_{lmn},
\end{eqnarray}
and the other components vanish. 

It is known \cite{CFMP} that the condition $\beta^{\cal G}_{MN}=\beta^{\cal B}_{MN}=0$ ensures the existence of a 
Virasoro algebra with central charge $c=6\beta^\Phi$. 
Since $\beta^{\cal G}_{mn}=0,\beta^{\cal B}_{mn}=0$ coincide with the equations of motion of $G_{mn},B_{mn}$, 
respectively, one can take a background 
\begin{eqnarray}
&& g_{\mu\nu} = \eta_{\mu\nu}, \hspace{5mm} B_{\mu\nu} = 0, \hspace{5mm} \Phi=\mbox{const.},  \nonumber \\
&& \phi^I=\mbox{const. \hspace{3mm}such that} \hspace{3mm} \frac{\partial V}{\partial \phi^I} = 0, 
   \label{solution}
\end{eqnarray}
and obtains the central charge 
\begin{eqnarray}
c \ \ = \ \ 6\beta^\Phi &=& -\frac1{4}G^{il}G^{jm}G^{kn}H_{ijk}H_{lmn} \nonumber \\
&=& \frac32V. 
    \label{c=V}
\end{eqnarray}

\vspace{3mm}

Suppose that the scalar potential $V$ has two critical points $\phi^I=\phi^I_1,\phi^I_2$ with 
$V(\phi_1)>V(\phi_2)$, and $\phi^I=\phi^I_1$ corresponds to an unstable vacuum. 
If $V(\phi_1)=0$, then (\ref{solution}) is indeed a classical solution and it defines a perturbative vacuum with 
vanishing central charge. 
This vacuum would be expected to decay into the second vacuum $\phi^I=\phi^I_2$, but since $V(\phi_2)<0$, the 
corresponding classical solution would be a linear dilaton background and worldsheet analyses in this background 
would not be reliable due to string loop corrections. 
What we can say at least is that the configuration (\ref{solution}) with $\phi^I=\phi^I_2$ would be in the 
vicinity of the linear dilaton background in the theory space and the central charge for the configuration is 
smaller than the initial central charge, due to (\ref{c=V}). 

This result would be interpreted as follows. 
The transition from (\ref{solution}) with $\phi^I_1$ to that with $\phi^I_2$ would correspond to a condensation 
of a bulk tachyon. 
It is natural since this is a shift of vevs of the scalars, and the central charge of the sigma model decreases. 
The transition from (\ref{solution}) with $\phi^I_2$ to the linear dilaton background would be a backreaction 
of the condensation. 
During this process, the dilaton behaves non-trivially and the worldsheet analyses would not valid. 
Note that the real process would not be able to devided into such two periods, and the direct investigation of 
such a simultaneous process would be difficult. 

If $V(\phi_2)=0$, the endpoint of the decay is well understood. 
However, in this case the initial unstable vacuum would be strongly-coupled, by assumption, and the investigation 
of whole process of the decay would be also difficult. 
 
To summarize, it seem to be good to apply the RG prescription to bulk tachyon condensations, but it might not be 
so useful since the backreaction would change the background drastically.

\vspace{1cm}

\section{Rolling down with constant dilaton}  \label{Dindep}

\vspace{5mm}

As we briefly commented in \cite{suyama}, when the scalar potential does not depend on the dilaton, transitions 
between critical points of the potential would occur without going into a strongly-coupled string theory. 
Then the same puzzle discussed in the beginning of the previous section will reappear in these cases.  
Thus we have to investigate such cases and find what is going on. 

We would like to consider a string theory whose low energy effective theory contains a non-trivial scalar 
potential which is a function independent of the dilaton $\Phi$. 
There is a supergravity theory of this kind \cite{5dim1}\cite{5dim2} which is familiar in the study of AdS/CFT 
correspondence \cite{AdS/CFT}. 
This is a five-dimensional theory with maximal supersymmetry, which is believed to be a consistent truncation of 
Type IIB supergravity compactified on $S^5$. 
It contains 42 scalars which is identified with the coordinates of the coset $E_{6(6)}/USp(4)$, and the scalar 
potential is a function on the coset. 
The potential is invariant under $SL(2,{\bf R})\subset E_{6(6)}$, and this corresponds to the 
$SL(2,{\bf R})$ invariance of Type IIB supergravity. 
Thus the potential is independent of the dilaton. 
Note that we have discussed in the Einstein frame. 
In the string frame, the scalar potential would be multiplied by an exponential of $\Phi$. 

The scalar potential has many critical points \cite{5dim1}\cite{critical1}\cite{critical2}\cite{critical3} 
and they are interpreted, in AdS/CFT correspondence, IR fixed 
points of the ${\cal N}$=4 super Yang-Mills theory in four dimensions perturbed by relevant deformations. 
Then, according to the correspondence, transitions between the critical points would also be relevant 
deformations of the corresponding Type IIB string theory. 
Therefore AdS/CFT correspondence would imply that there may be such a process which connects two critical string 
theories corresponding to the critical points, although along the whole process the tree level approximation 
seems to be reliable. 

Since the relation between the five-dimensional supergravity mentioned above and a sigma model is not clear, 
the investigation of this case is not possible. 
However, the problem is more general and it can be stated as follows. 
If $\Phi=$const. is a classical solution regardless of the value of the scalar potential, then tachyon 
condensations could occur without going into a strong coupling background and the RG prescription for tachyon 
condensations would not apply to this case. 
Below we will argue that the situation is quite different from the ones discussed in the previous section and 
the comparison of central charges would be a subtle issue. 

\vspace{3mm}

What we have done in the previous section is to consider background field configurations satisfying 
\begin{eqnarray}
\beta^{\cal G}_{MN} &=& {\cal R}_{MN}+2\hat{\nabla}_M\hat{\nabla}_N\Phi-\frac14{\cal H}_{MKL}{{\cal H}_N}^{KL}
                \ \  = \ \ 0, \\
\beta^{\cal B}_{MN} &=&-\frac12\hat{\nabla}^L{\cal H}_{LMN}+\hat{\nabla}^L\Phi\ {\cal H}_{LMN}
                 \ \ = \ \ 0, \\
\Phi &=& \mbox{const.} 
   \label{constantPhi}
\end{eqnarray}
The configurations are not necessarily on-shell configurations since the last condition (\ref{constantPhi}) is not 
equivalent to $\beta^\Phi=0$. 

When we further require that (\ref{constantPhi}) satisfies the equation of motion of $\Phi$, the allowed 
configurations are only on-shell configurations. 
One can see this from the fact that the equation of motion of $\Phi$ can be rewritten as 
\begin{eqnarray}
0 &=& {\cal R}+4\hat{\nabla}^2\Phi-4(\hat{\nabla}\Phi)^2-\frac1{12}{\cal H}^2 \nonumber \\
&=& {\cal G}^{MN}\beta^{\cal G}_{MN}-4\beta^\Phi,
\end{eqnarray}
i.e., the equation of motion of $\Phi$ is equivalent to $\beta^\Phi=0$ provided that $\beta^{\cal G}_{MN}=0$. 

Let us reconsider central charges in this case. 
Since for every critical point of the scalar potential one has to take an on-shell solution in order to ensure 
the existence of the Virasoro algebra, one cannot obtain, in this approach, a sigma model with a non-zero central 
charge. 
Therefore it is tempting to conclude that in this case central charge does not change in the decay process. 

However, there is a difference from the cases in the previous section. 
In those cases, the background geometry can be always taken to be flat, but in this case, since the background 
has to be on-shell, the spacetime is curved according to the value of the scalar potential. 
Thus one has to compare central charges for different background geometries. 
Recall that the central charge is a measure of the degrees of freedom whose definition strongly depends on the 
background geometry. 
Thus a naive comparison would not make sense. 
In the next section, we will argue a relation between the central charge of sigma models and the energy of the 
system in the spacetime sense, and a similar subtlety will appear there.

\vspace{1cm}

\section{Central charge as spacetime energy} \label{center}

\vspace{5mm}

We have discussed the central charge of a sigma model in a special situation in which all but one beta-functional 
for the background fields vanish. 
There is an expression of the central charge, or c-function, for a more general situation \cite{Tseytlin}. 

The Zamolodchikov's c-function $C(r)$ is defined in terms of correlation functions of a two-dimensional field 
theory, 
\begin{equation}
C(r) = 2F(r)-G(r)-\frac38H(r),
    \label{c-fn}
\end{equation}
where $r^2=z\bar{z}$ and 
\begin{eqnarray}
F(r) &=& z^4\langle T_{zz}(z,\bar{z})T_{zz}(0,0)\rangle, \nonumber \\
G(r) &=& 4z^3\bar{z}\langle T_{zz}(z,\bar{z})T_{z\bar{z}}(0,0)\rangle \nonumber \\
     &=& 4z^3\bar{z}\langle T_{z\bar{z}}(z,\bar{z})T_{zz}(0,0)\rangle, 
          \label{c_fn}\\
H(r) &=& 16z^2\bar{z}^2\langle T_{z\bar{z}}(z,\bar{z})T_{z\bar{z}}(0,0)\rangle\ge 0. \nonumber 
\end{eqnarray}
We have assumed the rotational invariance. 
Then the derivative of $C(r)$ is non-positive
\begin{equation}
r\frac d{dr}C(r) = -\frac34H(r),
\end{equation}
due to the conservation of the energy-momentum tensor. 

The sigma model considered in \cite{Tseytlin} corresponds to the ``spatial part'' of the sigma model 
describing a string theory. 
Thus the background fields of the sigma model consist of $G_{mn},B_{mn},\Phi$ in our notation. 
The c-function (\ref{c-fn}) is given in terms of the fields, and it is shown that the derivative of the 
c-function is non-positive and the vanishing of the derivative means that the sigma model is conformal. 
Explicitly, the c-function is given as 
\begin{eqnarray}
C(r) &=& \int_0^\infty \frac{dk}kf(rk)\cdot 2\kappa^2S(k), 
   \label{c=S} \\
S(k) &=& \frac1{2\kappa^2}\int d^{10-d}x\sqrt{G}e^{-2\Phi}\left[ 
         R-4\nabla_m\Phi\nabla^m\Phi+4\nabla^m\nabla_m\Phi-\frac1{12}H_{lmn}H^{lmn} \right].
\end{eqnarray}
Here $f(x)$ is a function which is necessary for the Fourier transformation from $k$ to $r$. 
By integrating by parts, it is shown that $S(k)$ is a part of the low energy effective action. 
The dependence on $k$, the renormalization scale, of $S(k)$ comes from that of the fields themselves. 

The equation (\ref{c=S}) relates the c-function to the effective action. 
This relation is natural for several reasons. 
Firstly, since both the conformal invariance condition and the on-shell condition can be characterized by the 
extremization of c-function and the effective action, respectively, the equivalence of these two conditions 
would imply the relation like (\ref{c=S}). 
It would be more apparent when the RG equations are written as follows, 
\begin{equation}
k\frac d{dk}G_{mn} = \frac{2\kappa^2}{\sqrt{G}}e^{2\Phi}\left[ 
                     \frac{\delta S}{\delta G^{mn}}+\frac14G_{mn}\frac{\delta S}{\delta \Phi} \right], 
                     \hspace{5mm} \mbox{etc.}
\end{equation}
Secondly, under the condition $\beta^G_{mn}=0$, the action can be written as 
\begin{equation}
S(k) = \frac1{2\kappa^2}\int d^{10-d}x\sqrt{G}e^{-2\Phi}(-4)\beta^\Phi.
\end{equation}
Therefore it is consistent with the fact that the central charge is proportional to $\beta^\Phi$ under the 
conditions discussed previously. 

It is well-known that the value of the Zamolodchikov's c-function at a conformal fixed point is equal to the 
central charge of the conformal theory. 
For the c-function (\ref{c=S}), however, it is not the case. 
This is because in \cite{Tseytlin} the c-function is derived from correlators which are not normalized as usual, 
and in particular, 
\begin{equation}
\langle 1\rangle = V(k) \equiv \int d^{10-d}x\sqrt{G}e^{-2\Phi}.
\end{equation}
Then the central charge should be related to $S(k)/V(k)$, and it need not be a monotonic function due to the 
scale dependent factor $V(k)$. 
In other words, the decreasing quantity is the c-function, not the central charge. 
Moreover, when $V(k)\to\infty$, the central charge does not change even when the c-function decreases by a finite 
amount. 
This fact would explain why the central charge does not decrease during condensations of localized closed string 
tachyons \cite{HKMM}. 

\vspace{3mm}

The natural extension of (\ref{c=S}) to the c-function of a sigma model for whole spacetime would be to replace 
$S(k)$ with the total spacetime effective action $S_{total}(k)$. 
It is known that the spacetime action is equal to the time integral of the Hamiltonian (times $-1$) under some 
conditions \cite{GR}. 
Therefore, as discussed in \cite{GHMS}, the c-function is related to the energy of the system in the spacetime 
sense. 
From this point of view, the meaning of the c-theorem is very clear; the physical process occurs toward a 
state with lower energy than the initial one. 
One can also see from this relation that the comparison of central charges for different backgrounds is a subtle 
issue; this corresponds to comparing energies defined in different background geometries. 

When the dilaton has a non-trivial configuration, 
however, one has to be careful not to conclude immediately that decreasing the c-function corresponds to 
decreasing the energy. 
It is because the effective action related to the c-function is written in the string frame, while the action 
related to the energy is in the Einstein frame. 
Their difference is a surface term, and it is irrelevant for the constant dilaton case \cite{GHMS} but otherwise 
would change the value of the action and the relation between the c-function and the energy may break down. 
This would be another explanation why the RG prescription for tachyon condensation would not valid for analysis 
of whole process of the condensation.

\vspace{1cm}

\section{Internal space geometry from scalar vevs}  \label{potential}

\vspace{5mm}

In the previous section, we have discussed the relation between the c-function of a sigma model and the action of 
the low energy effective theory, which can be related to the total spacetime energy of the system. 
This relation would provide a clear understanding of the c-theorem; the energetically-favored process occurs 
physically. 
Moreover, a slight difference between the c-function obtained in \cite{Tseytlin} and the total energy would 
suggest that there might be some modification of the c-theorem when the dilaton becomes non-trivially, that is, 
the tree level approximation for string theory is not reliable. 

In the rest of this paper, we will show another use of the spacetime theory to investigate a condensation of a 
closed string tachyon propagating in the bulk. 
That is, we deduce a final state of the tachyon condensation. 

\vspace{3mm}

We consider in this section a gauged supergravity which is a consistent truncation of a string theory or 
M-theory compactified on a non-trivial manifold. 
The theory has a scalar potential, which has many critical points. 
Some of them correspond to stable vacua of the theory, and the others to unstable vacua. 
When one defines the theory on one of the unstable vacua, it would be deformed via a condensation of an unstable 
mode, until the theory will reach another vacuum which is then stable. 
Therefore, one can obtain the knowledge of the condensation by examining the shape of the scalar potential. 

Recall that the scalars in the gauged supergravity in lower dimensions originally come, in principle, from fields 
of a theory in higher dimensions, 
for example, metric of the internal manifold. 
Thus the vevs of the scalars in a lower dimensional theory should have information of the target space geometry in 
a higher dimensional theory. 
This is indeed the case, and some of the critical points of the scalar potential are related to some internal 
manifolds \cite{AdS/CFT}. 
Moreover, in a particular case, there is a general formula \cite{embedding} to obtain the metric of the internal 
manifold from the vevs of the scalars. 
It is clear that bulk tachyon condensations inevitably deform background geometries, since they cannot be 
decoupled from the bulk gravity. 
Therefore it is important to know how the background geometry is deformed via the tachyon condensation. 
The relation between the scalar vevs and the internal geometry will enable one to investigate it. 

The most interesting final state of the condensation would be the most stable vacuum. 
In ordinary supersymmetric field theory without gravity, the most stable vacua is the states with zero energy. 
This is also the case for open string tachyon condensations in superstring \cite{open}. 
However, in supergravity theories the situation is different since its scalar potential is not necessarily 
non-negative. 
Moreover, in many cases the potential is not even bounded below. 
Naively, in such cases, the system would roll down the potential eternally toward the ``bottom". 
Thus it would be interesting to ask to what kind of theory it is deformed by such an evolution of the scalar vevs. 
We discuss this issue below with explicit examples.

\vspace{5mm}

\subsection{M-theory on $S^7$}

\vspace{5mm}

In this subsection, we discuss an ${\cal N}=8$ gauged supergravity in four dimensions \cite{CJ}, which is a 
consistent truncation of the eleven-dimensional supergravity compactified on $S^7$ \cite{truncation}. 
This theory has been well-studied and several critical points of its scalar potential have been found 
\cite{extrema}. 

$S^7$ can be regarded as an $S^1$-bundle over $CP^3$. 
Therefore, M-theory compactified on $S^7$ can be regarded as Type IIA string theory compactified on $CP^3$ with 
R-R fluxes. 
From this, we can regard instabilities which appear in the gauged supergravity as unstable modes of the Type IIA 
strings. 
Since the eleven-dimensional metric of the classical solution is a direct product $AdS_4\times S^7$, the R-R 
fluxes do not depend on the non-compact ($AdS$) coordinates. 
If there is an unstable mode coming from scalars, it is a tachyon in the bulk and thus its condensation would 
decrease the effective c-function discussed in section \ref{center}. 
The situation considered is similar to the Melvin background in M-theory \cite{Melvin}, except for the 
compactness of the relevant geometry. 

\vspace{3mm}

The gauged supergravity contains 70 scalars which can be identified with the coordinates of the coset 
$E_{7(7)}/SU(8)$. 
It is convenient to represent them in terms of a group element of the $E_{7(7)}$ which acts on the {\bf 56} 
representation, 
\begin{equation}
{\cal V} = \left(
\begin{array}{cc}
{u_{ij}}^{IJ} & v_{ijKL} \\ \bar{v}^{klIJ} & {\bar{u}^{kl}}_{KL}
    \label{vielbein}
\end{array}
\right).
\end{equation}
We follow the notations and conventions of \cite{notation}. 
The indices take values from 1 to 8, and $ij$, $KL$ etc. are anti-symmetrized. 
Note that an $SU(8)$ part of ${\cal V}$ is not physical degrees of freedom. 

The scalar potential is given as 
\begin{equation}
{\cal P(V)} = \frac1{24}|A_2|^2-\frac34|A_1|^2,
   \label{P(V)}
\end{equation}
where the tensors ${A_1}^{ij}$ and ${{A_2}_i}^{jkl}$ are complicated functions of $u$ and $v$ in 
(\ref{vielbein}). 
One can see that the potential has a negative-definite term, and this potential is indeed unbounded below. 

The directions along which the ${\cal P(V)}$ diverges in a limit correspond to the non-compact directions of 
$E_{7(7)}$, or more conveniently, the non-compact directions of $SL(8,{\bf R})\subset E_{7(7)}$. 
Explicitly, we consider one-parameter subgroups of $E_{7(7)}$ corresponding to 
\begin{equation}
X = \left(
\begin{array}{cccc}
e^{a_1s} & & & \\ & e^{a_2s} & & \\ & & \ddots & \\ & & & e^{a_8s}
   \label{SL(8)}
\end{array}
\right) \hspace{5mm} \in \ SL(8,{\bf R}), 
\end{equation}
with $\sum a_k=0$. 
The explicit form of the potential (\ref{P(V)}) has been calculated in special cases \cite{explicitP}, 
by using as ${\cal V}$ the element of the one-parameter subgroup corresponding to (\ref{SL(8)}), 
\begin{eqnarray}
{\cal P}_{7,1} &=& \frac18(-35e^{2s}-14e^{-6s}+e^{-14s}), \\
{\cal P}_{6,2} &=& -3(e^{2s}+e^{-2s}), \\
{\cal P}_{5,3} &=& -\frac38(5e^{2s}+10e^{-2s/3}+e^{-10s/3}), \\
{\cal P}_{4,4} &=& -(e^{2s}+4+e^{-2s}), 
\end{eqnarray}
where ${\cal P}_{p,8-p}$ is the potential (\ref{P(V)}) for the element (\ref{SL(8)}) with 
\begin{equation}
a_1=\cdots=a_p, \ a_{p+1}=\cdots=a_8,
\end{equation}
and their values are suitably chosen. 
This corresponds to the direction which preserves $SO(p)\times SO(8-p)$ symmetry. 
Note that ${\cal P}_{p,q}(s)={\cal P}_{q,p}(-ps/q)$. 
A point $s=0$ is always a critical point which corresponds to the round $S^7$. 
One can see that there are directions along which ${\cal P}\to-\infty$ as $|s|\to\infty$. 

\vspace{3mm}

As mentioned above, the vevs of the scalars contain information of the deformation of $S^7$. 
How $S^7$ is deformed for $|s|\to\infty$ ?
To investigate it, one has to reconstruct the metric of the $S^7$ from the scalar vevs. 
This was achieved in \cite{embedding}, and in particular, the simple formula for the metric for the vevs 
corresponding to (\ref{SL(8)}) was obtained. 
The $S^7$ is deformed to an ellipsoid defined by 
\begin{equation}
x_mP_{mn}x_n = \mbox{const.}, \hspace{5mm} (x_1,\cdots,x_8) \in {\bf R}^8,
\end{equation}
where $P=X^2$, and the metric on this ellipsoid is the induced metric with a scale factor $\mu$,
\begin{eqnarray}
g_{mn} &=& 2^{2/9}\mu^{-2/3}(\delta_{mn}-\hat{n}_m\hat{n}_n), \\
\mu^2 &=& x_mP_{ml}P_{ln}x_n,
\end{eqnarray}
where $\hat{n}_m$ is the unit normal vector of the ellipsoid. 
From this result, one can see that the $x_k$-direction shrinks (extends) when $a_k>0$ ($a_k<0$) as 
$s\to +\infty$, when the overall factor is ignored. 
Since $X$ is an element of $SL(8,{\bf R})$, not all the directions simultaniously shrink or extend. 
In other words, there must be a direction which extends as the scalar vevs grow. 

Among the directions for ${\cal P}\to -\infty$, an interesting case would be the direction which preserves 
$SO(2)\times SO(6)$ symmetry. 
In this case, the deformed $S^7$ is given as 
\begin{equation}
e^{3t}((x_1)^2+(x_2)^2)+e^{-t}((x_3)^2+\cdots+(x_8)^2)=\rho^2
\end{equation}
in ${\bf R}^8$, and $t\propto s$. 
The metric on this ellipsoid which is manifestly $SO(2)\times SO(6)$ invariant is 
\begin{equation}
ds^2 = \rho^2\left[ \left(e^t+e^{-3t}\frac{r^2}{1-r^2}\right)dr^2+e^tr^2d\Omega_5^2+e^{-3t}(1-r^2)d\theta^2
       \right],
\end{equation}
where $0\le r\le1$, and we have used a parametrization
\begin{eqnarray}
&x_1 = e^{-3t/2}\rho\sqrt{1-r^2}\cos\theta,& \\
&x_2 = e^{-3t/2}\rho\sqrt{1-r^2}\sin\theta,& \\
&(x_3)^2+\cdots +(x_8)^2 = \rho^2e^tr^2,&
\end{eqnarray}
so that this is a product of a six-dimensional ball and a circle whose radius depends on $r$. 
When $t$ is large, then 
\begin{equation}
ds^2 \sim \rho^2\left[ d\tilde{r}^2+\tilde{r}^2d\Omega_5^2+e^{-3t}(1-e^{-t}\tilde{r}^2)d\theta^2\right],
\end{equation}
where $\tilde{r}=e^{t/2}r$. 
Therefore, in the large $t$ limit, the ellipsoid is approximately the direct product of the six-dimensional flat 
space and the small $S^1$ with a constant radius, as long as $\tilde{r}<<e^{t/2}$. 
By choosing this $S^1$ as the M-theory circle, this limiting theory would be regarded as a weak coupling 
Type IIA string theory. 
Thus this might suggest that a tachyon condensation of this theory would have a final state which is a 
weakly-coupled Type IIA string theory in the flat spacetime. 
Note that no R-R 1-form appears via the dimensional reduction. 
Since the $S^1$ chosen above is not the same circle with the one, reducing along which provides Type IIA string 
theory compactified on $CP^3$, the relation between the initial and the final states would be non-trivial. 

However, the argument given above is too naive since we did not take into account the overall factor $\mu$. 
In fact, the proper metric is, up to a trivial numerical factor,
\begin{eqnarray}
ds^2_M &=& \rho^2\mu^{-2/3}
    \left[ \left(e^t+e^{-3t}\frac{r^2}{1-r^2}\right)dr^2+e^tr^2d\Omega_5^2+e^{-3t}(1-r^2)d\theta^2\right], \\
\mu^2 &=& e^{3t}(1-r^2)+e^{-t}r^2.
\end{eqnarray}
Through the dimensional reduction along the $\theta$ direction, one obtains the metric for the corresponding 
Type IIA string theory
\begin{equation}
ds^2_{IIA} = \rho^2e^{-3t/2}\mu^{-1}\sqrt{1-r^2}
        \left[ \left(e^t+e^{-3t}\frac{r^2}{1-r^2}\right)dr^2+e^tr^2d\Omega_5^2 \right].
\end{equation}
Then, one can calculate the size of this space and show that 
\begin{equation}
\int_0^1dr\sqrt{g_{rr}} \to 0\hspace{5mm} (t\to \infty).
\end{equation}
Therefore, the internal space becomes small as the scalar vevs grow. 
Note that the radius of the M-theory circle we chose is small everywhere for large $t$, so that the resulting 
theory should be a weak coupling theory. 
However, the compactification manifold is a strange one, and the resulting theory would not be familiar for us.


\vspace{3mm}

A similar phenomenon will occur in M-theory compactified on $S^4$, although the relation to string theory would 
be unclear. 
The consistent truncation of the eleven-dimensional supergravity is a seven-dimensional supergravity \cite{7dim}. 
The explicit form of the scalar potential along $SO(2)\times SO(3)$-invariant direction was obtained in 
\cite{7dimpotential}, and 
the internal metric was constructed \cite{embed7dim} in terms of the seven-dimensional fields.

\vspace{1cm}

\subsection{Type IIB string theory on $S^5$}

\vspace{5mm}

Since the analysis in this case is parallel to that in the previous subsection, we will discuss briefly. 

The truncated theory is believed to be the five-dimensional gauged supergravity mentioned before, 
and critical points of its 
scalar potential is 
well-studied. 
In fact, some of them are shown to be unstable \cite{critical1}\cite{critical3}. 
Such an unstable vacuum would decay via a tachyon condensation, and as a result, the scalar vev would roll 
down the potential toward its ``bottom''. 

As before, we focus on the direction which is invariant under $SO(2)\times SO(4)$. 
The explicit expression of the potential restricted to this direction parametrized be $\lambda$ is \cite{5dim1} 
\begin{equation}
{\cal P}(\lambda) = -\frac14(e^{2\lambda}+2e^{-\lambda}). 
\end{equation}
Note that the critical point $\lambda=0$ is the maximally supersymmetric vacuum, so that it cannot decay into 
$|\lambda|\to\infty$. 
However, some other unstable vacua might decay into a direction toward $|\lambda|\to\infty$. 

The formula for the metric of the internal $S^5$ in terms of the five-dimensional fields was conjectured in 
\cite{critical3}, 
and in particular, along the $SO(2)\times SO(4)$-invariant direction the $S^5$ is expected to be deformed 
to an ellipsoid
\begin{equation}
e^{-2t}((x_1)^2+(x_2)^2)+e^t((x_3)^2+\cdots+(x_6)^2)=\rho^2.
\end{equation}
Therefore, based on the same analysis in the previous subsection, this background will approach a strange 
manifold which is an $S^1$-bundle and the radius is small. 
Since we have considered Type IIB string theory, this final theory could be regarded as Type IIA string theory 
compactified on a small four-dimensional manifold.

\vspace{1cm}

\section{Discussion} \label{discussion}

\vspace{5mm}

We have investigated the condensations of the bulk closed string tachyons by using the spacetime effective 
theory. 
It would be the most suitable situation in which the effective theory is a gauged supergravity, and then the 
tachyons appear when one choose an unstable vacuum. 
The spacetime supersymmetry strongly restrict the dynamics, so that it should be expected that one could control 
part of quantum corrections, even when one would like to analyze a non-supersymmetric background. 

\vspace{3mm}

We have considered only the cases in which the gauge group is compact. 
This is because one can relate the vevs of the scalars to the geometry of the internal manifold. 
There are many other gauged supergravities with non-compact gauge groups, and critical points of the scalar 
potentials were found \cite{explicitP}\cite{5dim1}. 
However, their geometric origins do not seem to be understood well until now. 
It is expected that relations between such gauged supergravities and higher dimensional theories would provide us 
more examples of bulk tachyon condensations, and our understanding of the condensations would become deeper. 
A proposed correspondence \cite{DW/QFT} will be relevant in this direction of research. 

\vspace{3mm}

Another generalization can be considered. 
Since we have focused on the maximally supersymmetric theories, the scalar potentials are completely determined 
by the supersymmetry. 
It is plausible since then we can obtain explicit formulae. 
But if we would like to have a more general potential to consider a more general situation, we have to reduce 
the number of supersymmetry. 
There is a nice characterization of a kind of ${\cal N}=2$ gauged supergravities in four dimensions. 
They can be realized as the low energy effective theory of Type II string theory compactified on a manifold with 
the SU(3) structure (see e.g. \cite{SU(3)str} and references theirin). 
A six-dimensional manifold $M$ with the SU(3) structure admits a single SU(3) invariant spinor on $M$ which is 
not required to be covariantly constant. 
For this reason, the four-dimensional Minkowski spacetime 
is not, in general, a classical solution of the effective 
theory, that is, there is a non-trivial scalar potential. 
This realization of gauged supergravities will provide various examples of theories with unstable vacua, and 
one could extract a general feature of the decays of them. 
The investigation of a meaning of the SU(3) structure in terms of a two-dimensional theory is also interesting, 
and would be important to understand the mirror symmetry in the presence of flux. 

\vspace{3mm}

We have argued that the sigma model approach would not apply to whole process of the bulk tachyon condensations. 
However, it would be still useful to study properties of each vacuum. 
It is well-known that the notion of stability depends on what background the theory is defined \cite{BFbound}. 
To understand them in terms of the sigma model would be interesting. 

\vspace{3mm}

Our interest has been mainly on the bulk tachyon condensations. 
The usefulness of the spacetime action might, however, continue to hold for the localized tachyon condensations 
\cite{localized}. 
Localized tachyons usually come from twisted sectors, and thus it would be realized by just adding the 
corresponding matter fields in the effective action. 
Then it would be possible to do the similar analysis as we have done in this paper. 

\vspace{2cm}

{\bf {\Large Acknowledgements}}

\vspace{5mm}

We would like to thank T. Kimura, N. Sakai, T. Takayanagi and Y. Yasui for valuable discussions.

\end{document}